\documentclass[PAPER, cernpreprint, atlasdraft=false, UKenglish, texmf, orcidlogo]{atlasdoc}
\usepackage{atlaspackage}
\usepackage{multirow}
\usepackage{atlasbiblatex}

\usepackage{atlasphysics}
\usepackage{atlassnippets}
\graphicspath{{logos/}{Figures/}}

\usepackage{ANA-HMBS-2024-25-PAPER-defs}

\AtlasTitle{Search for Higgs boson exotic decays into Lorentz-boosted light bosons in the four-$\tau$ final state at $\sqrt{s} = 13\,\mathrm{TeV}$ with the ATLAS detector}

\AtlasAbstract{%

A search for exotic decays of the Higgs boson into a pair of low-mass scalars that subsequently decay into $\tau$-leptons, $H\rightarrow aa\rightarrow \tau^+\tau^-\tau^+\tau^-$, is presented. In models with Yukawa-like couplings, the decay to $\tau$-leptons is favoured for light $a$-bosons, with mass in the range of $2m_{\tau} < m_a < 2m_{b}$. Results are presented in the range of $4\,\mathrm{GeV} < m_a < 15\,\mathrm{GeV}$ using the $140\,\mathrm{fb}^{-1}$ of proton--proton collisions at $\sqrt{s} = 13\,\mathrm{TeV}$ recorded with the ATLAS detector during $\mathrm{Run}\,2$ of the Large Hadron Collider. This search focuses on the scenario where, for both di-$\tau$ pairs, one of the $\tau$-leptons decays to hadrons and neutrinos, while the other decays to a muon and neutrinos. In this mass range, the $a\rightarrow \tau^+\tau^-$ is Lorentz-boosted and a dedicated muon removal technique is used to reconstruct the di-$\tau$ pairs. No significant excess above the Standard Model background prediction is observed. Upper limits on $(\sigma(H)/\sigma_{\text{SM}}(H))\times \mathcal{B}(H\rightarrow aa\rightarrow 4\tau)$ at $95\%$ confidence level are provided, ranging from 0.03 to 0.10 depending on the $a$-boson mass.

}

\AtlasRefCode{HMBS-2024-25}

\PreprintIdNumber{CERN-EP-2025-038}

\AtlasJournal{Phys.\ Lett.\ B.}
\AtlasJournalRef{\PLB 870 (2025) 139843}
\AtlasDOI{10.1016/j.physletb.2025.139843}

\AtlasCoverEgroupAnalysisTeam{atlas-hmbs-2024-25-analysis-team@cern.ch}

\hypersetup{pdftitle={ATLAS document},pdfauthor={The ATLAS Collaboration}}

\begin{document}

\maketitle

\tableofcontents

\section{Introduction}
%

%
%
%
Following the observation of the Higgs boson by the ATLAS~\cite{HIGG-2012-27} and CMS~\cite{CMS-HIG-12-028} collaborations at the Large Hadron Collider (LHC)~\cite{Evans:2008zzb}, both experiments are conducting an extensive programme to measure its properties and uncover its fundamental nature. Measurements of the Higgs boson couplings to other Standard Model (SM) particles can constrain the branching ratio to beyond-the-SM (BSM) particles to be less than 12\% at a 95\% confidence level (CL), assuming that the Higgs boson's coupling to vector bosons is not larger than in the SM ($\kappa_V\leq 1$)~\cite{HIGG-2021-23,CMS-HIG-22-001}.

Exotic Higgs boson decays to BSM particles are proposed as a way to detect new physics that is weakly coupled to the SM.\@ Due to the narrow total width of the Higgs boson in the SM, even a small coupling to a new light state could induce a significant change in BSM branching ratio~\cite{Curtin:2013fra,Cepeda:2021rql}. Furthermore, new particles might preferentially couple to the Higgs boson, making it a potential \textit{portal} for hidden-sector particles to interact with SM particles~\cite{Strassler:2006im, Schabinger:2005ei, Patt:2006fw}. Exotic Higgs boson decays are possible in models of dark matter~\cite{Silveira:1985rk, POSPELOV200853, Draper:2010ew},  electroweak baryogenesis~\cite{Carena:2022qpf,Kozaczuk:2019pet}, and neutral naturalness~\cite{Craig:2015pha, Curtin:2015fna}. They are also predicted from first principles in the next-to-minimal supersymmetric SM~\cite{Dobrescu:2000yn, Ellwanger:2003jt}.

In models where the SM is extended by a light scalar, $a$, if the $a$-boson can mix with the Higgs boson it will inherit its Yukawa coupling pattern to fermions. In such cases, the $a$-boson will decay preferentially to the heaviest kinematically allowed fermion pair. Although the $a$-boson is referred to as a scalar in this paper, the results presented are equally applicable to both the scalar and pseudoscalar cases. For light $a$-bosons, in the mass range $2m_{\tau} < m_a < 2m_{b}$, one of the preferred decays is $a\to \tau^+\tau^-$. Due to the low mass of the $a$-boson, this decay is Lorentz-boosted and the two $\tau$-leptons overlap in the detector. This search focuses on the case in which, for each $a$-boson, one of the $\tau$-leptons decays to a muon $\tau\to \mu\nu_{\mu}\nu_{\tau}$ and the other decays to hadrons $\tau\to \text{hadrons}+\nu_{\tau}$, referred to as \tauhad. In this decay mode, the muon can be identified close to the hadrons and, once removed from the hadron reconstruction cone, it does not impact the identification of the hadronically decaying $\tau$-lepton. This paper presents a search for events with two di-$\tau$ pairs identified through a muon-removal technique, consistent with a decay $H\to aa\to \tau^+\tau^-\tau^+\tau^-$ in the mass range $4\,\GeV<m_a<15\,\GeV$. When $m_a$ is higher, the two $\tau$-leptons do not overlap, requiring a different analysis strategy.

The CMS Collaboration has previously searched for exotic Higgs boson decays $H\to aa\to \tau^+\tau^-\tau^+\tau^-$ using only tracks to identify the hadronically decaying $\tau$-leptons~\cite{CMS-HIG-18-006}. The ATLAS Collaboration has also searched for exotic Higgs boson decays in the $H\to aa\to \mu^+\mu^-\tau^+\tau^-$ decay channel using only tracks to identify the $\tau$-lepton decays~\cite{HIGG-2014-02}. The muon-removal technique allows an improved detection of the hadronically decaying $\tau$-leptons using both tracks and calorimeter information, as described in Ref.~\cite{TAUP-2023-02} and used for the first time by the ATLAS Collaboration in this paper. A similar muon-removal reconstruction has also been used by the CMS Collaboration in the search for exotic Higgs boson decays in the $H\to aa\to \mu^+\mu^-\tau^+\tau^-$ decay channel~\cite{CMS-HIG-18-024}.

This paper complements many previous searches for exotic Higgs boson decays $H\to aa$ conducted by the ATLAS and CMS Collaborations. In addition to the results mentioned above, searches are performed in the $\ell^+\ell^-\ell^+\ell^-$\cite{HDBS-2018-55, CMS-HIG-19-007} (where $\ell$ is an electron or a muon), $b\bar{b}\tau^+\tau^-$~\cite{HDBS-2021-07, CMS-HIG-22-007}, $b\bar{b}\mu^+\mu^-$~\cite{HDBS-2021-03,CMS-HIG-18-011}, $b\bar{b}b\bar{b}$~\cite{HDBS-2018-47, CMS-HIG-18-026}, $\gamma\gamma\gamma\gamma$~\cite{HDBS-2019-19,CMS-HIG-21-016,CMS-HIG-21-003}, and $\gamma\gamma gg$~\cite{HIGG-2017-09} final states.


%

%
\section{ATLAS detector}
%

%
%
%
\newcommand{\AtlasCoordFootnote}{%
ATLAS uses a right-handed coordinate system with its origin at the nominal interaction point (IP)
in the centre of the detector and the \(z\)-axis along the beam pipe.
The \(x\)-axis points from the IP to the centre of the LHC ring,
and the \(y\)-axis points upwards.
Polar coordinates \((r,\phi)\) are used in the transverse plane,
\(\phi\) being the azimuthal angle around the \(z\)-axis.
The pseudorapidity is defined in terms of the polar angle \(\theta\) as \(\eta = -\ln \tan(\theta/2)\) and is equal to the rapidity
$ y = \frac{1}{2} \ln \left( \frac{E + p_z c}{E - p_z c} \right) $ in the relativistic limit.
Angular distance is measured in units of \(\Delta R \equiv \sqrt{(\Delta y)^{2} + (\Delta\phi)^{2}}\).}

The ATLAS experiment~\cite{PERF-2007-01} at the LHC is a multipurpose particle detector
with a forward--backward symmetric cylindrical geometry and a near \(4\pi\) coverage in
solid angle.\footnote{\AtlasCoordFootnote}
It consists of an inner tracking detector surrounded by a thin superconducting solenoid
providing a \qty{2}{\tesla} axial magnetic field, electromagnetic and hadronic calorimeters, and a muon spectrometer.
The inner tracking detector covers the pseudorapidity range \(|\eta| < 2.5\).
It consists of silicon pixel, silicon microstrip, and transition radiation tracking detectors.
Lead/liquid-argon (LAr) sampling calorimeters provide electromagnetic (EM) energy measurements
with high granularity within the region \(|\eta|< 3.2\).
A steel/scintillator-tile hadronic calorimeter covers the central pseudorapidity range (\(|\eta| < 1.7\)).
The endcap and forward regions are instrumented with LAr calorimeters
for EM and hadronic energy measurements up to \(|\eta| = 4.9\).
The muon spectrometer surrounds the calorimeters and is based on
three large superconducting air-core toroidal magnets with eight coils each.
The field integral of the toroids ranges between \num{2.0} and \qty{6.0}{\tesla\metre}
across most of the detector.
The muon spectrometer includes a system of precision tracking chambers up to \(|\eta| = 2.7\) and fast detectors for triggering up to \(|\eta| = 2.4\).
The luminosity is measured mainly by the LUCID--2~\cite{LUCID2} detector, which is located close to the beampipe.
A two-level trigger system is used to select events~\cite{TRIG-2016-01}.
The first-level trigger is implemented in hardware and uses a subset of the detector information
to accept events at a rate below \qty{100}{\kHz}.
This is followed by a software-based trigger that
reduces the accepted event rate to \qty{1}{\kHz} on average
depending on the data-taking conditions.
A software suite~\cite{SOFT-2022-02} is used in data simulation, in the reconstruction
and analysis of real and simulated data, in detector operations, and in the trigger and data acquisition
systems of the experiment.


%

%
%

%
\section{Data and simulated event samples}
%

%
%
%
This search uses the full LHC \RunTwo data of proton--proton ($pp$) collisions at $\sqrt{s}=13$~\TeV recorded with the ATLAS detector from 2015 to 2018, corresponding to an integrated luminosity of $140.1 \pm 1.2$~fb$^{-1}$~\cite{DAPR-2021-01}. Only data taken from stable beam collisions and that satisfy a standard set of data-quality requirements that ensure all the ATLAS subdetectors were functioning correctly are considered~\cite{DAPR-2018-01}.

The data are selected from a suite of triggers requiring one or two high-transverse-momentum ($\pt$) muons~\cite{TRIG-2018-01}. Since the muons in the signal of interest here are produced close to hadrons, only triggers without isolation requirements are used. The lowest \pt threshold in triggers requiring a single non-isolated muon is $50\,\GeV$. The fraction of selected signal events satisfying this trigger requirement varies from 19\% to 42\% for $m_a$ from 4~\GeV to 15~\GeV. During the 2016--2018 (2015) data-taking period, the triggers requiring two muons have either a symmetric threshold of $\pt > 14\,\GeV$ ($\pt > 10\,\GeV$) for both muons or an asymmetric threshold that requires the leading muon to have $\pt > 22\,\GeV$ ($\pt > 18\,\GeV$) and the subleading muon to have $\pt > 8\,\GeV$ ($\pt > 8\,\GeV$). The efficiency of the di-muon trigger varies from 35\% to 43\% over the mass range from 4~\GeV to 15~\GeV.

Monte Carlo (MC) samples are used to perform the analysis optimisation, to estimate the signal acceptance and efficiency, and to describe background processes with four $\tau$-leptons in the final state. Signal samples $H\to aa\to \tau^+\tau^-\tau^+\tau^-$ are generated via gluon--gluon fusion (ggF) at next-to-leading-order (NLO) accuracy in quantum chromodynamics (QCD) using \POWHEGBOX[v2]~\cite{Hamilton:2013fea,Hamilton:2015nsa,Alioli:2010xd,Nason:2004rx,Frixione:2007vw} with the \NNPDF[3.0nnlo] parton distribution function (PDF) set~\cite{Butterworth:2015oua}. The signal generation achieves next-to-next-to-leading-order (NNLO) accuracy for inclusive ggF observables by reweighting the Higgs boson rapidity spectrum in Higgs plus one jet multi-scale improved NLO (\textsc{Hj}-\MINLO)~\cite{Hamilton:2012np,Hamilton:2012rf} to that of Higgs NNLO (HNNLO)~\cite{Bozzi:2005wk}. The Higgs boson decay into two scalar $a$-bosons and the subsequent decay of each $a$-boson into a pair of $\tau$-leptons is generated with \PYTHIA v8.245~\cite{Sjostrand:2014zea}. Signal samples are generated with $m_a$ = 4, 6, 8, 10, 12, 14, and 15~\GeV. Although only the ggF production mode is explicitly generated, the difference in acceptance compared to other production modes is negligible, and the sample is used to represent inclusive Higgs boson production.

The main source of background events is processes with fewer than four $\tau$-leptons in the final state, but in which one or two jets originating from the hadronisation of quarks or gluons are misidentified as hadronically decaying $\tau$-leptons. These processes are collectively called \textit{fake-$\tauhad$ background} hereafter and they are estimated with a data-driven technique described in \Sect{\ref{sec:background}}.

Background events with four $\tau$-leptons in the final state are subleading and described with simulation events. Only three processes are found to have a contribution greater than 0.1\% of the total background and thus considered relevant: non-resonant $\qqbar\to ZZ$ and $gg\to ZZ$ production, and resonant Higgs boson $H\to ZZ^*$ production. The non-resonant $\qqbar\to ZZ$ process is modelled using continuum quark--antiquark annihilation from \SHERPA[2.2.2]~\cite{Bothmann:2019yzt}, which provides a matrix element calculation accurate at NLO in QCD for 0-jet and 1-jet final states and at leading-order (LO) for 2-jet and 3-jet final states. The merging with the \SHERPA parton shower~\cite{Schumann:2007mg} is performed using the MEPS@NLO prescription~\cite{Hoeche:2012yf}. The loop-induced non-resonant $gg\to ZZ$ process is also modelled by \SHERPA[2.2.2] with 0-jet and 1-jet final states at LO in QCD.\@ The LO-accurate matrix elements are matched to a parton shower using the MEPS@LO prescription. For all non-resonant $ZZ$ processes modelled using \SHERPA, the \NNPDF[3.0nnlo] PDF set~\cite{Butterworth:2015oua} is used, along with a dedicated \AZNLO set of tuned parton-shower parameters~\cite{STDM-2012-23}.

Two different production mechanisms are generated for the resonant $H\to ZZ^*$ background. The ggF and vector-boson-fusion (VBF) production modes are generated at NLO accuracy with \POWHEGBOX[v2] using the CT10 PDF set~\cite{Lai:2010vv}. Other production modes with smaller cross-sections do not contribute significantly and are not included. The Higgs boson decay, showering, and hadronisation are generated with \PYTHIA v8.212 using the \AZNLO parameter set.

All generated events are processed through a simulation of the ATLAS detector geometry and response using \GEANT~\cite{Agostinelli:2002hh}, and through the same reconstruction software as the data.
The signal MC samples are processed with a fast simulation that relies on a parameterisation of the calorimeter response~\cite{SOFT-2010-01}.
The effect of multiple interactions in the same and neighbouring bunch crossings (\pileup) was modelled by overlaying the simulated hard-scattering event with inelastic \pp events generated with \PYTHIA[8.186]~\cite{Sjostrand:2007gs} using the \NNPDF[2.3lo] PDF set~\cite{Ball:2012cx} and the A3 set of tuned parameters~\cite{ATL-PHYS-PUB-2016-017}.


%

%
%

%
\section{Object and event selection}
\label{sec:selection}

%
%
%
The signal object selected is a $\mu\tauhad$ object, defined as an overlapping \tauhad and a muon with $\Delta R(\mu, \tau_{\text{had}})< 0.4$ and opposite charges. The $\mu\tau_{\text{had}}$ object is used to reconstruct the di-$\tau$ system in the $a\to \tau^+\tau^-$ decay where one $\tau$-lepton decays as $\tau\to \text{hadrons}+\nu_{\tau}$ and the other decays as $\tau\to \mu\nu_{\mu}\nu_{\tau}$ (charge conjugated processes are implicit).

Muons are reconstructed using the information from the inner-detector system, the muon spectrometer, and the calorimeters~\cite{MUON-2018-03}. Muon candidates are required to have $\pt > 5\,\GeV$ and $|\eta|< 2.5$. Muons must also satisfy the \textit{medium identification working point} defined in detail in Ref.~\cite{MUON-2018-03}. Muons from $\tau$-lepton decays can be displaced from the primary vertex due to the $\tau$-lepton lifetime, so a loose requirement on the transverse impact parameter significance $|d_0/\sigma_{d_0}| < 7$ is imposed. The longitudinal impact parameter is required to satisfy $|z_{0}\sin\theta|<0.5\,\text{mm}$ relative to the position of the primary vertex.

Electrons are reconstructed by matching clusters in the electromagnetic calorimeter to tracks in the inner detector~\cite{EGAM-2018-01}. Electron candidates are required to meet the criteria of $\pt > 7\,\GeV$ and $|\eta| < 2.47$, excluding the calorimeter transition region $1.37 < |\eta| < 1.52$. Additionally, all electrons must satisfy a \textit{loose} likelihood identification criterion~\cite{EGAM-2018-01}. %

The reconstruction of the \tauhad\ candidates starts from jets formed using the \antikt\ algorithm~\cite{Cacciari:2008gp} as implemented in \Fastjet~\cite{Fastjet} with a radius parameter $R=0.4$. This jet is referred to as seed jet, $\tau_{\text{seed}}$ jet. The inputs to the \antikt\ algorithm are topological clusters of calorimeter cells~\cite{PERF-2014-07}. The tracks associated with the $\tau_{\text{seed}}$ jet are used to determine the number of charged hadrons resulting from the decay of the $\tau$-lepton. To distinguish \tauhad candidates originating from hadronic $\tau$-lepton decays from those originating from jets initiated by the hadronisation of quarks or gluons, a recurrent neural network (RNN) identification algorithm~\cite{ATL-PHYS-PUB-2019-033} is used. The RNN uses information from reconstructed tracks and calorimeter energy clusters associated with the \tauhad candidates, as well as several high-level discriminating variables as input. A dedicated boosted decision tree (BDT) electron veto is also constructed to reject the backgrounds arising from electrons faking \tauhad~\cite{ATL-PHYS-PUB-2015-045}.

\begin{figure}[tbp!]
\centering
\subfloat[]{
\includegraphics[width=0.48\textwidth]{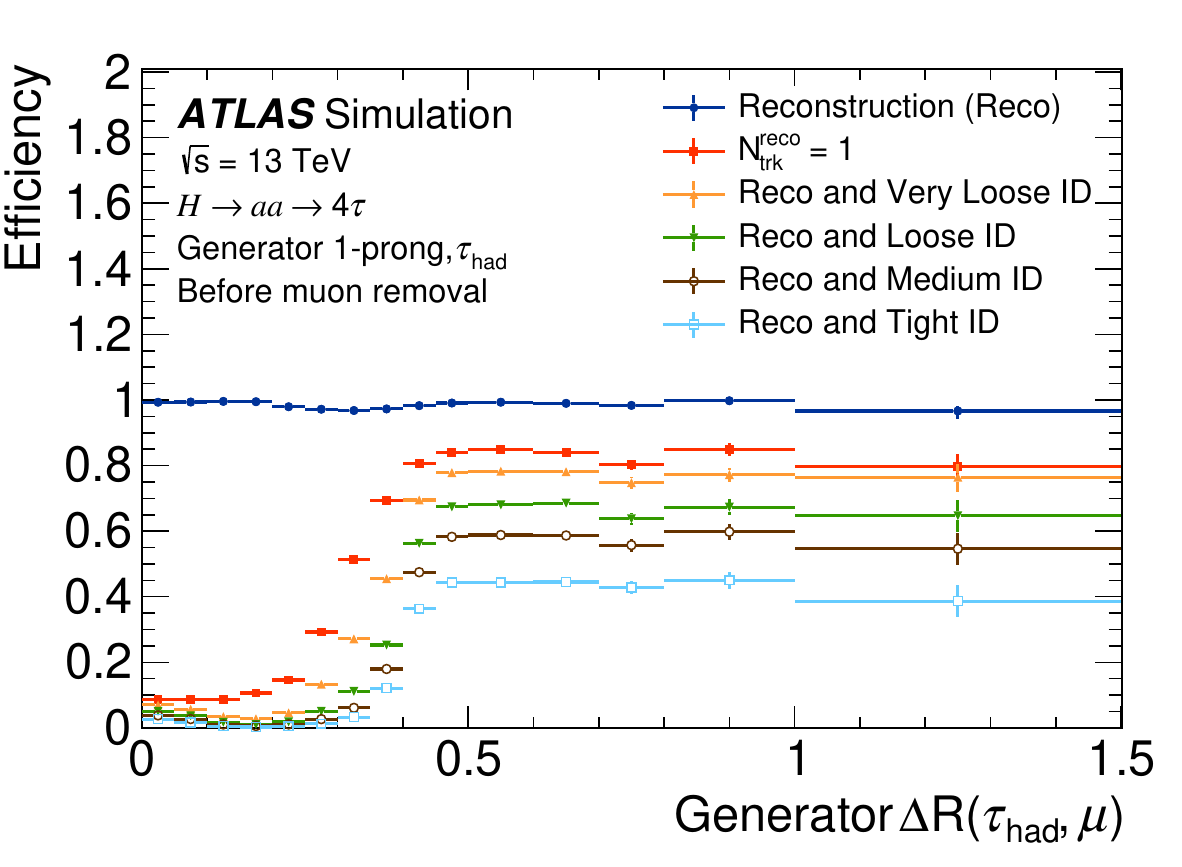}
\label{fig:RecoIDEff1prongStdTaus}
}\hfill
\subfloat[]{
\includegraphics[width=0.48\textwidth]{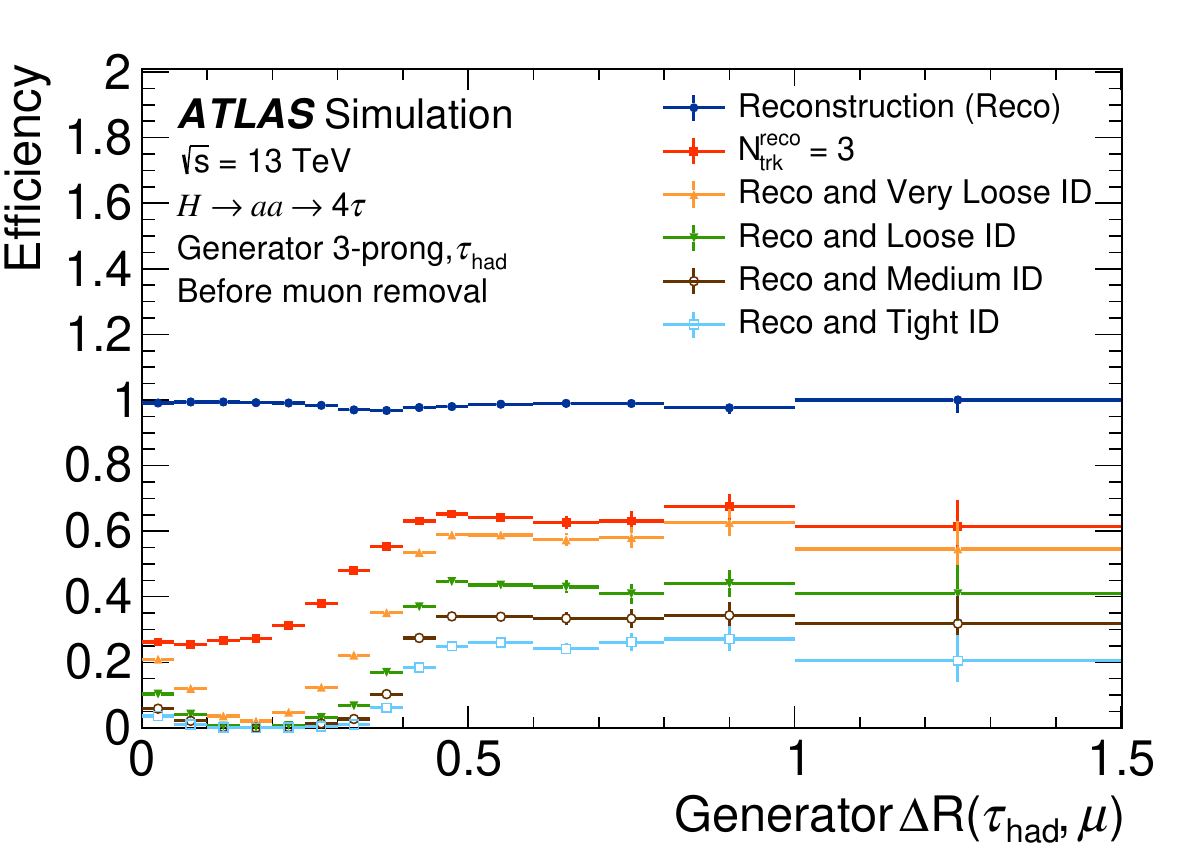}
\label{fig:RecoIDEff3prongStdTaus}
}\\
\subfloat[]{
\includegraphics[width=0.48\textwidth]{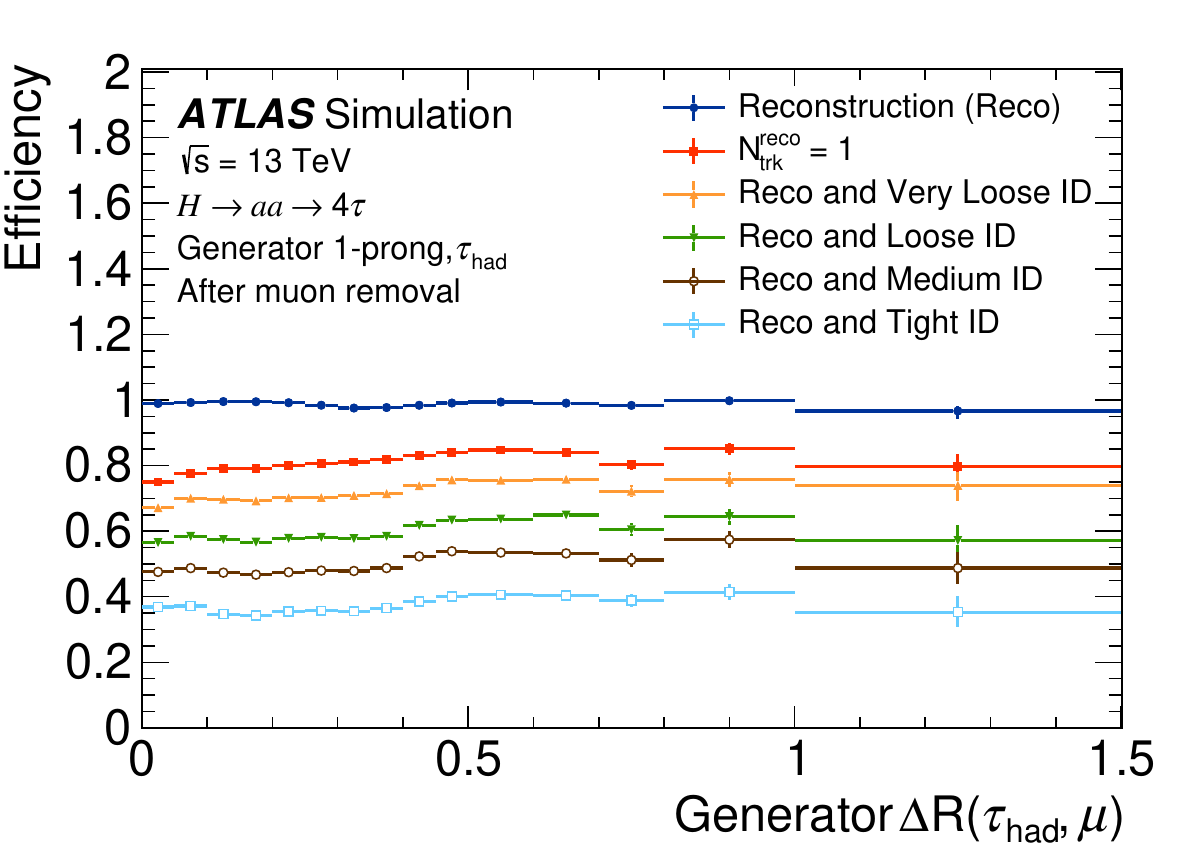}
\label{fig:RecoIDEff1prongMRTaus}
}\hfill
\subfloat[]{
\includegraphics[width=0.48\textwidth]{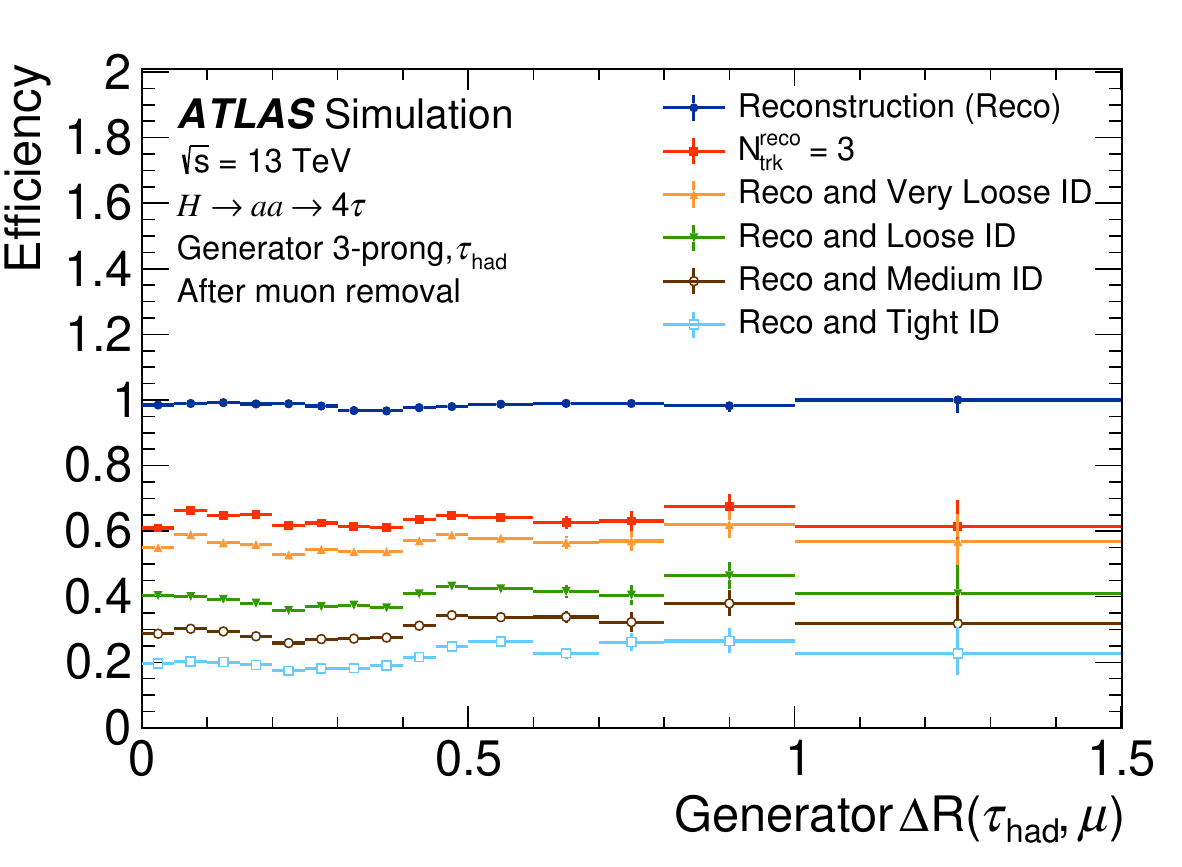}
\label{fig:RecoIDEff3prongMRTaus}
}
\caption{
Monte Carlo estimates of the $\tauhad$ reconstruction and RNN identification efficiency (\protect\subref{fig:RecoIDEff1prongStdTaus} and \protect\subref{fig:RecoIDEff3prongStdTaus}) before and (\protect\subref{fig:RecoIDEff1prongMRTaus} and \protect\subref{fig:RecoIDEff3prongMRTaus}) after the muon removal for generator-level (\protect\subref{fig:RecoIDEff1prongStdTaus} and \protect\subref{fig:RecoIDEff1prongMRTaus}) 1-prong and (\protect\subref{fig:RecoIDEff3prongStdTaus} and \protect\subref{fig:RecoIDEff3prongMRTaus}) 3-prong \tauhad candidates for all working points (\text{Very Loose}, \text{Loose}, \text{Medium}, \text{Tight}) defined in Ref.~\cite{ATL-PHYS-PUB-2019-033}, as a function of generator-level $\Delta R(\tauhad, \mu)$. The `Reconstruction (Reco)' markers indicate the efficiency of a $\tau_{\text{seed}}$ jet to match a generator-level \tauhad; The `$N_{\text{trk}}^{\text{reco}}=$1 or 3' markers show the efficiency of a generator-level \tauhad reconstructed with the same number of associated charged-particle tracks as charged hadrons at generator-level. These plots are based on a mixture of $m_a=4, 6, 8, 10, 15$~\GeV samples, with equal weighting applied to each sample. The slight decrease in reconstruction efficiency for both 1-prong and 3-prong \tauhad candidates within the range $0.2<\Delta R(\tauhad,\mu)<0.4$ is attributed to the less than 100\% generator-level matching for the signal samples with $m_a = 10$~\GeV and $15$~\GeV. This is a result of the strict geometrical matching criteria applied between the reconstructed \tauhad and the generator-level \tauhad.
\label{fig:TauPerformance}
}

\end{figure}

In $\mu\tau_{\text{had}}$ objects, the presence of the muon inside the $\tau_{\text{seed}}$ jet reduces the \tauhad\ reconstruction and RNN identification efficiency, as shown in Figures~\ref{fig:RecoIDEff1prongStdTaus} and \ref{fig:RecoIDEff3prongStdTaus} for $\tau$-lepton decaying to final states with one (1-prong) and three (3-prong) charged hadrons, respectively. To improve the reconstruction and identification efficiency of the \tauhad and thus of the $\mu\tau_{\text{had}}$ objects, a muon-removal technique, as described in Ref.~\cite{TAUP-2023-02}, is employed. In this muon-removal technique, the tracks and calorimeter energy clusters associated with the reconstructed muons within $\Delta R(\tauhad, \mu) < 0.4$ are removed from the inputs used for \tauhad\ reconstruction and from the inputs to the RNN and BDT identification algorithms.

The muon-removal technique is able to recover \tauhad reconstruction and RNN identification efficiencies that are similar to those of a \tauhad without a nearby muon. Figure~\ref{fig:TauPerformance} shows estimates of the reconstruction and RNN identification efficiency in simulation before and after applying the muon-removal technique for generator-level 1-prong and 3-prong \tauhad candidates for all the working points defined in Ref.~\cite{ATL-PHYS-PUB-2019-033}.
The efficiency is plotted as a function of the generator-level $\Delta R(\tauhad, \mu)$, allowing a distinction between merged \tauhad candidates ($\Delta R(\tauhad, \mu) < 0.4$) and isolated \tauhad candidates ($\Delta R(\tauhad, \mu) > 0.4$). Figure~\ref{fig:RecoIDEff1prongStdTaus} and Figure~\ref{fig:RecoIDEff3prongStdTaus} show the loss in RNN identification efficiency of the standard algorithm when the muon and the $\tauhad$ candidates overlap for 1-prong and 3-prong $\tauhad$ candidates, respectively. After applying the muon-removal technique, the efficiency values for 1-prong and 3-prong \tauhad candidates are shown in Figures~\ref{fig:RecoIDEff1prongMRTaus} and~\ref{fig:RecoIDEff3prongMRTaus}, respectively, demonstrating that this technique effectively allows to maintain the same efficiency for the merged and isolated cases. The `$N_{\text{trk}}^{\text{reco}} = 1$ or $3$' lines represent the efficiency of reconstructing a generator-level \tauhad with the same number of associated charged-particle tracks as charged hadrons at generator-level. A slight decrease in this reconstruction efficiency for $\Delta R(\tauhad, \mu) > 0.4$ after muon removal compared with the values before muon removal arises due to a small fraction of \tauhad charged hadrons being misidentified as muons and subsequently removed from the $\tau_{\text{seed}}$ jet.
The residual difference in \tauhad\ efficiency between isolated ($\Delta R(\tauhad,\mu)>0.4$) and merged ($\Delta R(\tauhad,\mu)<0.4$) $\tauhad$ candidates is 8\% for 1-prong and 10\% for 3-prong candidates, respectively. This difference is applied as a per-\tauhad\ uncertainty, as described in \Sect{\ref{sec:systematic}}.

The overlap of the muon with the $\tau_{\text{seed}}$ jet also impacts the \tauhad four-momentum and the $\tauhad$ energy scale. A MC-based correction using signal events is derived to correct the \tauhad four-momentum to the same energy scale as in isolated \tauhad. The effect of the correction on the background is also checked in regions rich in fake-$\tauhad$ background and is found to have a negligible impact on the agreement between data and background predictions. The uncorrected kinematic quantities, before the muon removal, are used for the event selection to minimize the dependence on simulation.

After the muon-removal technique is applied, the \tauhad\ candidates are required to satisfy uncorrected transverse momentum, $\pt^{\text{uncorr}} > 20\,\GeV$, and pseudorapidity criteria, $|\eta^{\text{uncorr}}| < 2.5$. Any candidate in the transition region between the barrel and endcap calorimeters, $1.37<|\eta^{\text{uncorr}}|<1.52$, is excluded. Two additional identification criteria are defined: \tauhad candidates satisfying the \textit{loose} identification criterion are required to have an RNN identification score above 0.01, which ensures a selection efficiency of 99\% for both 1-prong and 3-prong \tauhad candidates; \tauhad candidates satisfying the \textit{tight} identification criterion are required to satisfy the RNN working point that provides a signal efficiency of 75\% (60\%) for 1-prong (3-prong) $\tauhad$ candidates. All \tauhad candidates are required to satisfy the electron-veto BDT \textit{loose} working point with a signal efficiency of 95\% for both 1-prong and 3-prong $\tauhad$ candidates as defined in Ref.~\cite{ATL-PHYS-PUB-2015-045}.



An overlap removal procedure is applied on all objects (except the muons and the \tauhad) to eliminate double-counting as in Ref.~\cite{HIGG-2019-09}. The remaining objects are then used to perform the following event-level selection. The event selection is designed to reject events from most background sources, while maintaining a high acceptance to $H\to aa\to \tau^+\tau^-\tau^+\tau^-$ decays. Events are required to have exactly two $\mu\tauhad$ objects. The leading muon among the two objects is required to satisfy $\pt > 14\,\GeV$. The leading (subleading) \tauhad\ is required to satisfy $\pt^{\text{uncorr}} > 30\,\GeV$ ($\pt^{\text{uncorr}} > 25\,\GeV$) to reduce the number of background events with \textit{fake} $\tau$-leptons. Both \tauhad must satisfy the \textit{tight} identification criterion.

The uncorrected $\tauhad$ four-momenta and the muon four-momenta are used to reconstruct an estimate of the Higgs boson mass $m_{2\mu\tauhad}^{\text{uncorr}}$. Events are required to satisfy $60\,\GeV < m_{2\mu\tauhad}^{\text{uncorr}} < 120\,\GeV$. The selection is below the Higgs boson mass of 125~\GeV to account for the energy carried out by neutrinos. The corrected $\tauhad$ four-momenta are used to reconstruct an estimate of the two $a$-boson masses $m_{\mu\tauhad}$. Both $\mu\tauhad$ objects are required to satisfy $m_{\mu\tauhad} < 15\,\GeV$. The two mass-based selection criteria remove almost all the electroweak sources of background events while retaining approximately 90\% of the signal.

Two orthogonal signal regions (SR) are defined based on the relative charge of the two muons: a same-sign region (SS$\mu$), and an opposite-sign region (OS$\mu$). The fake-$\tauhad$ background in the OS$\mu$ region receives a large contribution from $Z$+jets events, which is absent in the SS$\mu$ region. An additional requirement $m_{\mu\mu}<50\,\GeV$ is imposed for events in the OS$\mu$ region to reduce this contribution and bring the background in both regions to similar levels while retaining 94\%~-~98\% of the remaining signal events depending on the $a$-boson mass.


%

%
\section{Background estimation}
\label{sec:background}

%
%
%
Background events from processes that have four $\tau$-leptons in the final state are modelled by simulation. Only non-resonant $\qqbar / gg\to ZZ$ and resonant $H\to ZZ^*$ processes are explicitly included in the analysis, but they have a negligible impact on the result. After event selection, these processes contribute to less than 0.1\% of the total background.

Almost all background events come from processes in which at least one jet from the hadronisation of a quark or a gluon is mis-reconstructed and mis-identified as a $\tauhad$ candidate (fake $\tauhad$). Simulation studies show that, in this case, the nearby muon is almost always from a non-prompt semileptonic decay of a hadron inside the jet. This background source is estimated with a tight-to-loose data-driven method~\cite{TAUP-2023-01} as briefly described below.

As described in \Sect{\ref{sec:selection}}, each $\tauhad$ candidate and, therefore each $\mu\tauhad$ candidate, has a \textit{loose} and a \textit{tight} identification criterion. Events with at least one $\mu\tauhad$ candidate that satisfies the \textit{loose} identification but not the \textit{tight} one are used to estimate the fake-$\tauhad$ background by assigning a per-$\mu\tauhad$ fake-factor weight $F$. The sign of the overall event weight is adjusted depending on the number of $\mu\tauhad$ failing to meet the \textit{tight} identification to avoid double-counting of background events~\cite{HDBS-2021-07,TAUP-2023-01}.

The fake factor $F$ is measured using $Z$+jets events selected by requiring two isolated muons and one $\mu\tauhad$ candidate satisfying the \textit{loose} identification requirement. The isolated muons are required to satisfy stricter impact parameter requirements $|d_0/\sigma_{d_0}| < 4$ and have an invariant mass close to the $Z$ boson pole mass $71\,\GeV < m_{\mu\mu} < 111\,\GeV$. The fraction of events $f$ in which the $\mu\tauhad$ also satisfies the \textit{tight} identification criterion is used to define the fake factor $F=f/(1-f)$, parameterised as a function of the $\mu\tauhad$ candidate $\pt^{\text{uncorr}}$ and the number of associated charged-particle tracks.

\begin{figure}[!tbp]
\centering
\subfloat[]{
\label{fig:mvismutauSSmuLinearPerbin}
\includegraphics[width=0.48\textwidth]{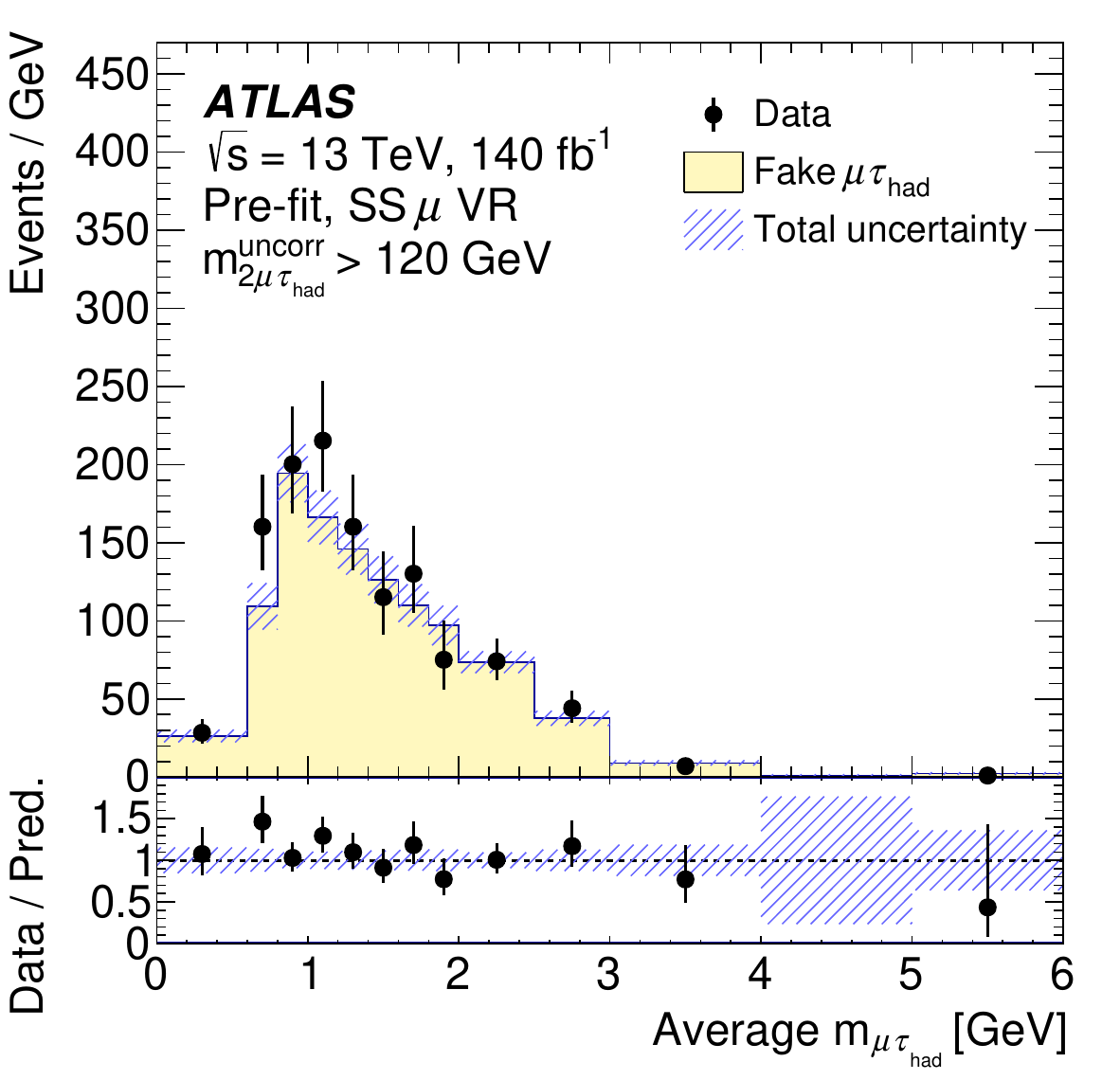}
}\hfill
\subfloat[]{
\label{fig:mvismutauOSmuLinearPerbin}
\includegraphics[width=0.48\textwidth]{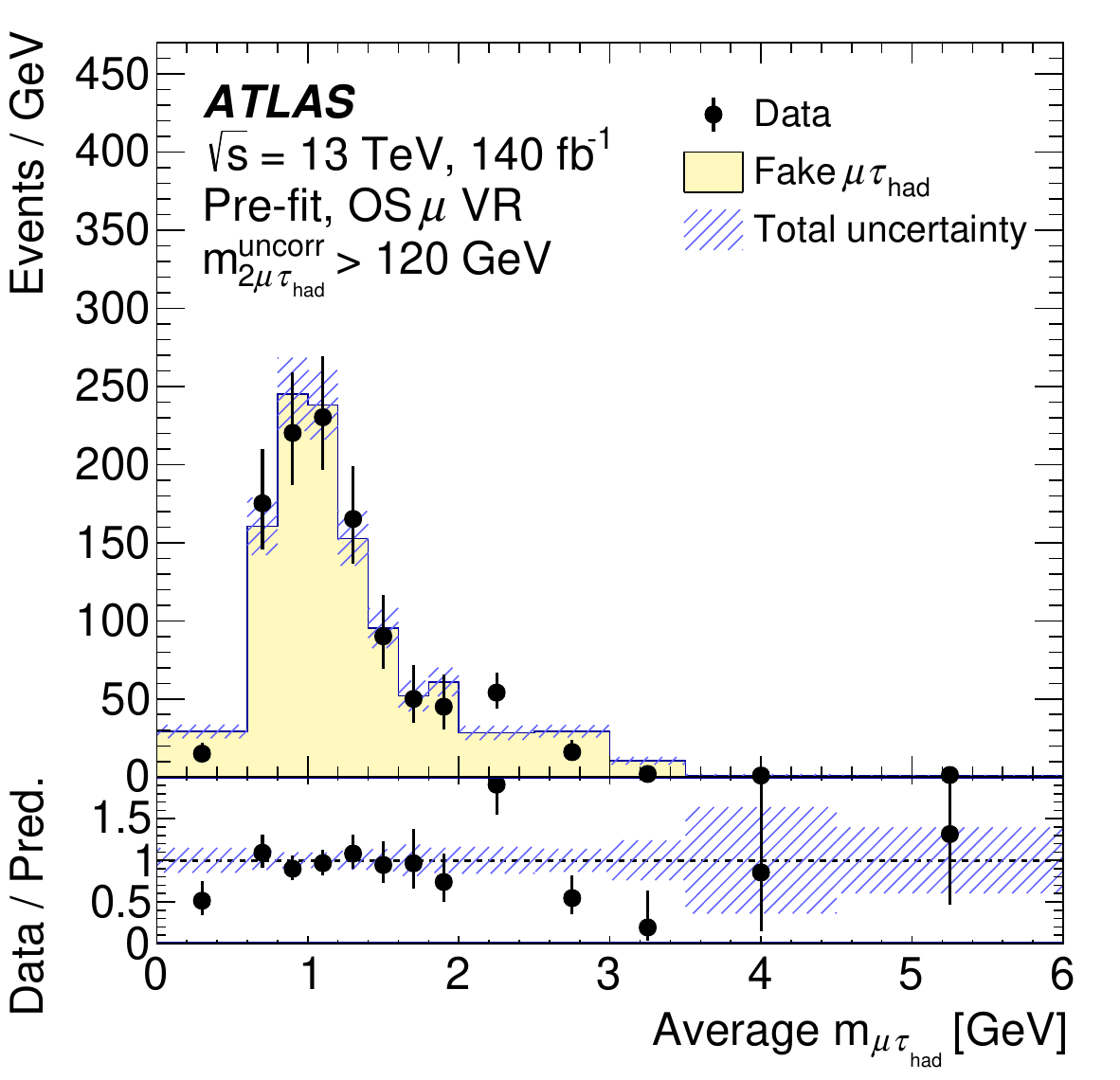}
}

\caption{
Pre-fit distribution of the average mass of the two $\mu\tauhad$ candidates in each event for data and the expected background in the \protect\subref{fig:mvismutauSSmuLinearPerbin} SS$\mu$ and \protect\subref{fig:mvismutauOSmuLinearPerbin} OS$\mu$ validation regions. The hashed area represents the total background uncertainty, including both statistical and systematic components. Overflow events up to 15~\GeV are included in the last bin. The contributions from $\qqbar/gg\to ZZ$ and $H\to ZZ^*$ are considered,  but they are not visible as they make up for less than 0.1\% in every bin.
}
\label{fig:HVMmvismutauLinearPerbin}
\end{figure}

The modelling of background events with fake $\mu\tauhad$ is verified in a validation region (VR), where events are selected in the same way as in the two signal regions, but with a high four-body invariant mass, $m_{2\mu\tauhad}^{\text{uncorr}} > 120\,\GeV$. Negligible signal contribution is expected in this region. The modelling is verified separately for events with same-charge (SS$\mu$ VR) and opposite-charge muon pairs (OS$\mu$ VR). \Fig{\ref{fig:HVMmvismutauLinearPerbin}} shows the average corrected mass of the two $\mu\tauhad$ candidates in each event of these validation regions for data and the expected background. This variable is used as a final discriminant in the signal regions, as explained in Section~\ref{sec:result}. The residual difference between data and expected background in this distribution is used as a non-closure systematic uncertainty, as described in \Sect{\ref{sec:systematic}}.

%


%

\section{Systematic uncertainties}
\label{sec:systematic}

Several sources of uncertainties in the signal acceptance are considered. Experimental sources include uncertainties in the muon reconstruction and identification efficiencies, in the muon momentum scale and in the muon trigger efficiency. The methods used to derive these uncertainties are described in Refs.~\cite{TRIG-2018-01,MUON-2018-03}. Similar systematic uncertainties related to the $\tauhad$ reconstruction, RNN identification, and BDT electron-veto efficiencies~\cite{ATL-PHYS-PUB-2019-033,ATLAS-CONF-2017-029} are considered. Uncertainties in the $\tauhad$ energy scale are included following the methods of Ref.~\cite{ATL-PHYS-PUB-2015-045}. An additional systematic uncertainty is included to account for the difference between identification efficiency in isolated $\tauhad$ candidates and $\mu\tauhad$ candidates after muon removal. The relative difference estimated in simulation between the two cases is 8\% (10\%) and applied as an uncertainty separately for each 1-prong (3-prong) $\tauhad$ candidate. Uncertainties in the simulation of \pileup interactions are considered, but have a negligible impact on the signal acceptance. The uncertainty in the integrated luminosity is 0.83\% \cite{DAPR-2021-01}, obtained using the LUCID-2 detector~\cite{LUCID2} for the primary luminosity measurements, complemented by measurements using the inner detector and calorimeters.

Modelling systematic uncertainties in the signal acceptance are estimated by varying the renormalisation and factorisation scales between half and twice the nominal value used in the simulation~\cite{Bothmann:2016nao}. Uncertainties related to the parton distribution function are estimated via \NNPDF MC replicas and by comparing the acceptance when different PDF sets~\cite{Butterworth:2015oua} are used.

The main sources of systematic uncertainty in this search are the ones related to the modelling of the fake-$\tauhad$ background. Uncertainties coming from the limited number of data events in the $Z$+jets region used to estimate the fake factor $F$ are considered but have a negligible impact on the result. Uncertainties coming from the limited number of data events with at least one $\mu\tauhad$ candidate failing to meet the tight identification in the two signal regions are more important, especially for the results at higher values of $m_a$ where they can be as large as 62\%. Variations related to the fake factor parameterisation on $\tauhad$ $\pt$ and $\eta$ are considered, but have a negligible impact on the result.

Several uncertainties are included to account for possible differences between the fake $\tauhad$ composition in the $Z$+jets region, where the fake factors are measured, and the signal regions, where they are applied. Alternative fake factors are measured using the same set-up and $Z$+jets region but requiring additional light jets, $b$-tagged jets, or restricting $\Delta R(\tauhad,\mu) < 0.2$. The differences between each of these three alternative fake factors and the nominal ones are considered as independent source of systematic uncertainty. Studies in simulation show that these three variations can account for the difference between fake-$\tauhad$ composition in the $Z$+jets region, signal regions, and validation regions. Finally, a non-closure uncertainty of 8\% in the fake-$\tauhad$ background total yield is estimated based on the relative difference between data and expected background in the SS$\mu$ and OS$\mu$ VRs.

Systematic uncertainties in the background processes modelled with simulation were studied but they have a negligible impact on the result since these background sources account for less than 0.1\% of the total background across all signal regions.


%

%
\section{Results}
\label{sec:result}

\begin{figure}[tbp!]
\centering
\subfloat[]{\label{fig:BonlyAvMvisSSmuLinearPerbin}
\includegraphics[width=0.48\textwidth]{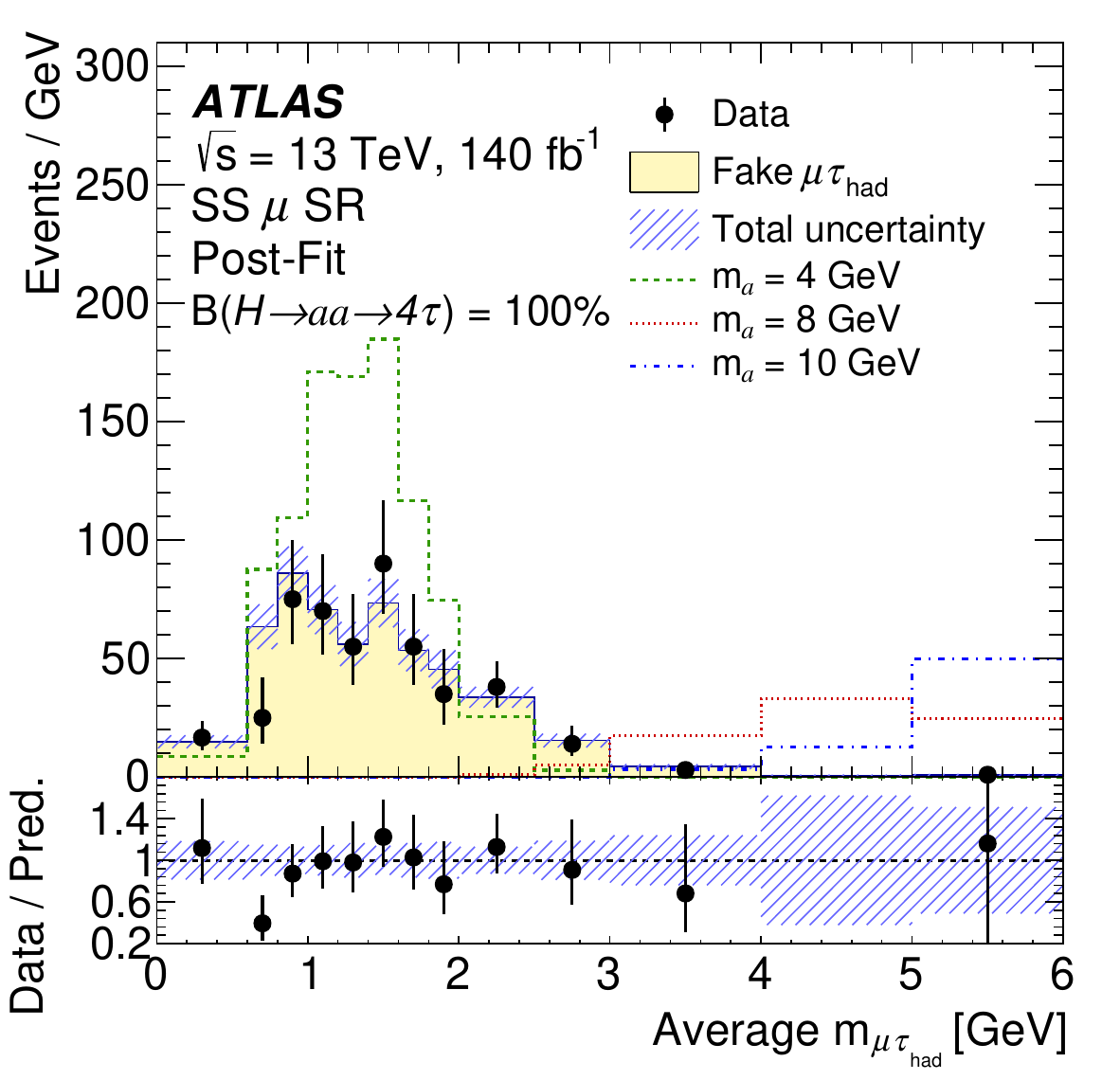}}\hfill
\subfloat[]{\label{fig:BonlyAvMvisOSmuLinearPerbin}
\includegraphics[width=0.48\textwidth]{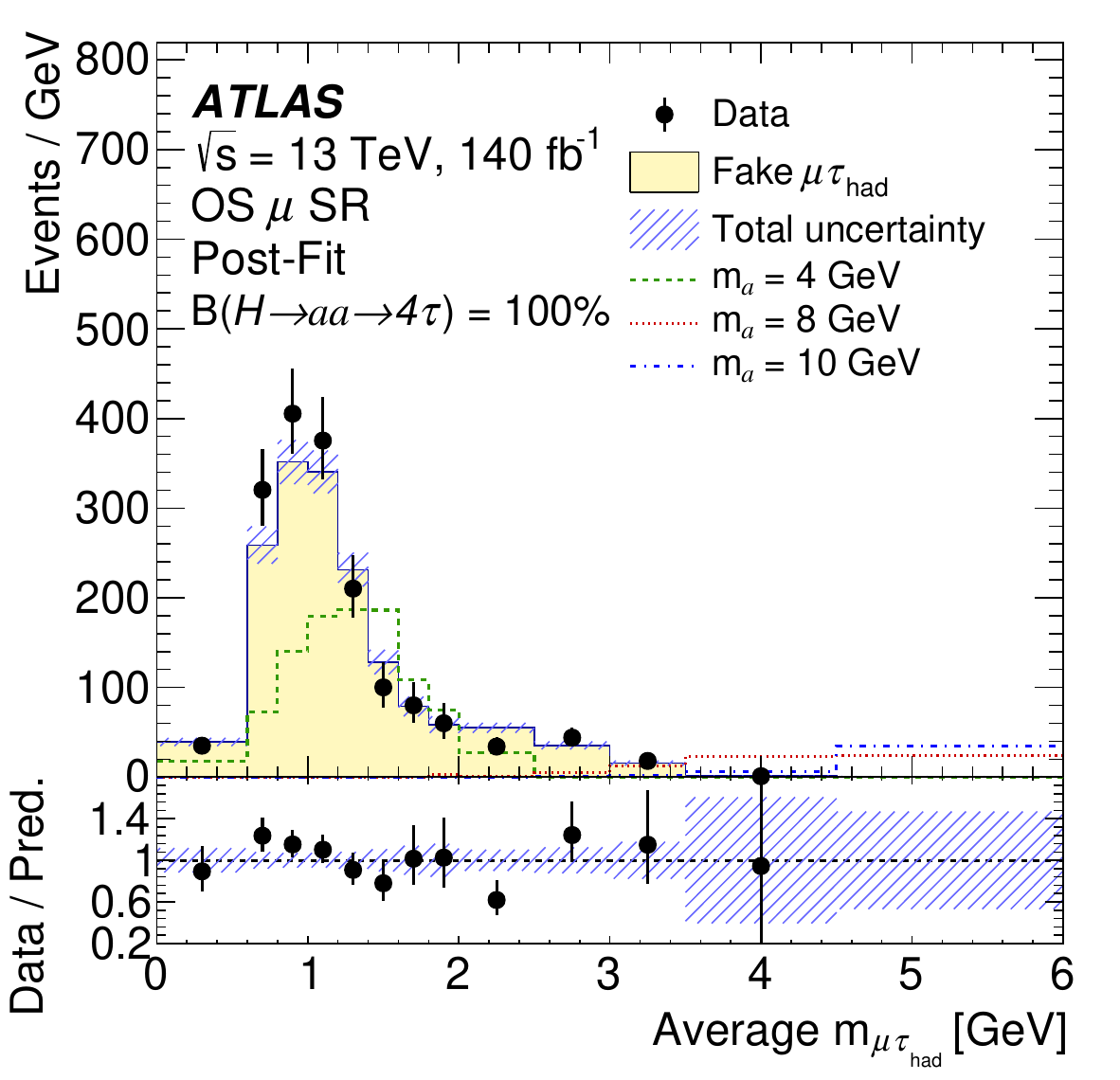}}
\caption{Post-fit distribution of the average mass of the two $\mu\tauhad$ candidates in each event for data, the expected background and $m_a=4$, 8, and 10~\GeV signals, assuming $\mathcal{B}(H\to aa\to 4\tau)=100\%$, in the \protect\subref{fig:BonlyAvMvisSSmuLinearPerbin} SS$\mu$ and \protect\subref{fig:BonlyAvMvisOSmuLinearPerbin} OS$\mu$ signal regions. Overflow events up to 15~\GeV are included in the last bin. The hashed area represents the total background uncertainty, including both statistical and systematic components. The contributions from $\qqbar/gg\to ZZ$ and $H\to ZZ^*$ are considered,  but they are not visible as they make up for less than 0.1\% in every bin. }
\label{fig:postfitSRsLinearPerbin}
\end{figure}

The average mass of the two $\mu\tauhad$ candidates in each event is used as a discriminating observable to test the presence of $H\to aa\to \tau^+\tau^-\tau^+\tau^-$ decays, due to its better resolution compared with the individual $\mu\tauhad$ masses.
The binning strategy for the SS$\mu$ and OS$\mu$ signal regions was optimised to target a statistical uncertainty of approximately 20\% in the prediction of background events, as depicted in Figure~\ref{fig:postfitSRsLinearPerbin}. The last two bins were specifically optimised to enhance the signal significance over all mass hypotheses simultaneously while ensuring non-zero background event yields in each bin. As a result, the statistical uncertainties in these bins exceed 20\%.

The presence of a $H\to aa\to \tau^+\tau^-\tau^+\tau^-$ decay is tested with a modified profile likelihood ratio $\tilde{q}(\mu)$~\cite{Cowan:2010js}, where $\mu = (\sigma(H)/\sigma_{\text{SM}}(H))\times \mathcal{B}(H\to aa\to 4\tau)$ and $\sigma_{\text{SM}}(H)$ is the inclusive Higgs boson cross-section at $\sqrt{s}=13\,\TeV$. The likelihood function is built as a product of Poisson probability functions for each average $m_{\mu\tauhad}$ bin in the SS$\mu$ and OS$\mu$ SRs. Nuisance parameters with Gaussian constraints are introduced to model sources of systematic uncertainties and their impact on the expected signal and background yields.

\begin{figure}[tbp!]
\centering
\includegraphics[width=0.55\textwidth]{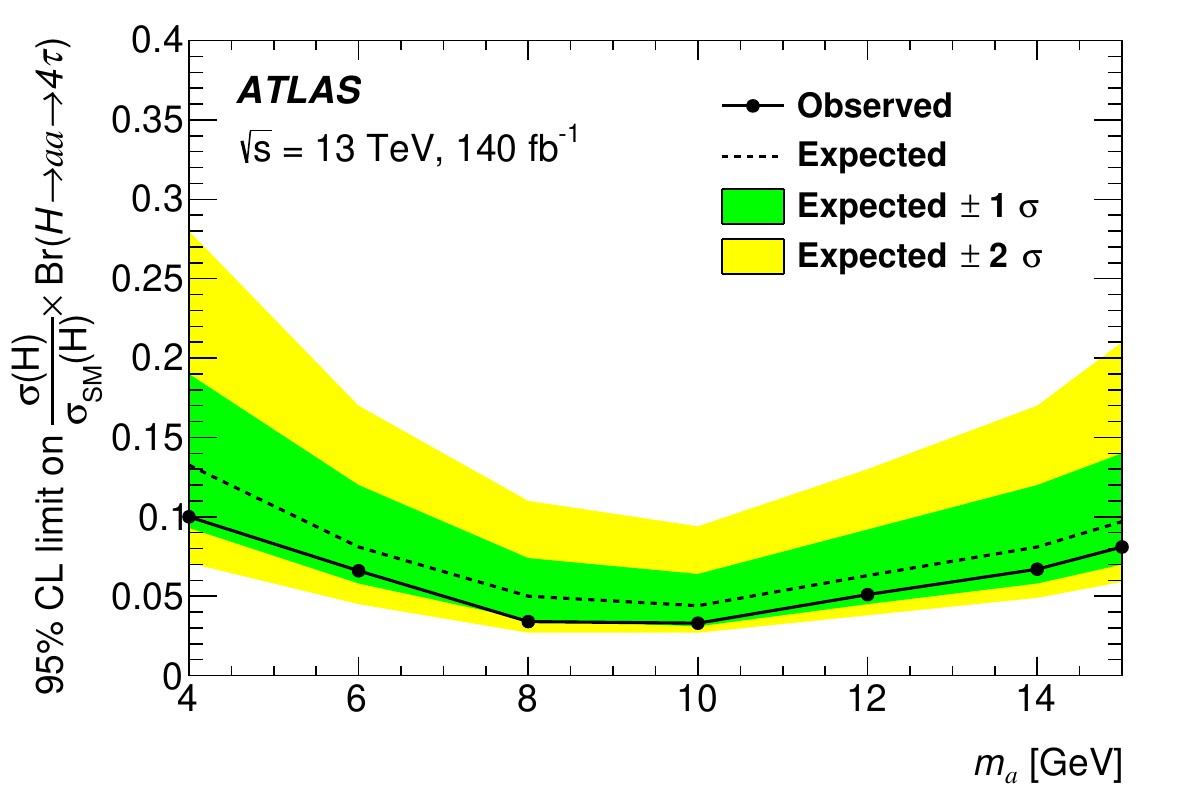}
\caption{Observed (solid line) and expected (dashed line) 95\% CL upper limits on $(\sigma(H)/\sigma_{\text{SM}}(H))\times\mathcal{B}(H\to aa\to 4\tau)$ as a function of $m_a$. The inner green and outer yellow shaded bands represent the $\pm 1$ and $\pm 2$ standard deviations around the expected limits, respectively. Limits between the mass points are interpolated linearly.\label{fig:Limits}}
\end{figure}

The observed number of events in the two SRs is shown in Table~\ref{tab:SRFinalYields} along with the predicted background.
The signal yields for different $m_a$ hypotheses are also shown assuming a $\mathcal{B}(H\to aa\to 4\tau)=10\%$ and the inclusive Higgs boson cross-section of $55.7\,\text{pb}$~\cite{deFlorian:2016spz}. No significant excess is observed in the SRs when comparing the total data yield to the total background prediction in the absence of a signal. The post-fit total number of background events is $130 \pm 7$ for the SS SR and $368 \pm 16$ for the OS SR. Upper limits are set at the 95\% CL on $(\sigma(H)/\sigma_{\text{SM}}(H))\times\mathcal{B}(H\to aa\to 4\tau)$ using the $\text{CL}_{\text{S}}$ method~\cite{Read:2002hq}, and assuming a reference cross section of $55.7\,\text{pb}$ for $\sigma_{\text{SM}}(H)$. Figure~\ref{fig:Limits} shows the observed and expected upper limits obtained using pseudo-experiments to estimate the test-statistic distribution for each value of the signal strength $\mu$ and $a$-boson mass hypothesis. The SS$\mu$ channel demonstrates greater sensitivity than the OS$\mu$ channel for $m_a<6$~\GeV due to less $Z/\gamma*\to \mu^+\mu^-$ background, whereas the SS$\mu$ and OS$\mu$ channels show similar sensitivities for $m_a \geq 6$~\GeV. The worse limits for $m_a < 8$~\GeV are due to the poor separation between the signal and background, as the signal shapes closely resemble those of the background. Additionally, the worse limits for $m_a > 10$~\GeV result from fewer merged $\tau$-leptons, leading to lower signal efficiency. The dominant uncertainties arise from the limited number of \pp collisions collected, ranging from 51\% to 70\%, depending on the mass of $a$-boson. The observed (expected) limit on $(\sigma(H)/\sigma_{\text{SM}}(H))\times\mathcal{B}(H\to aa\to 4\tau)$ ranges from 0.03 (0.04) at $m_a=10$~\GeV to 0.10 (0.13) at $m_a=4$~\GeV.

\begin{table}[tbp!]
\centering
\caption{The pre-fit table showing the total number of observed and expected events in the SS$\mu$ and OS$\mu$ signal regions. The uncertainties include both statistical and systematic sources. The expected number of signal events is listed for different $m_a$ hypotheses assuming $\mathcal{B}(H\to aa\to 4\tau)=10\%$ and the inclusive Higgs boson cross-section.}
\label{tab:SRFinalYields}%
\begin{tabular}{lrr}
\hline\hline\toprule
\textbf{Process} & \textbf{SS$\mu$ region} & \textbf{OS$\mu$ region} \\\midrule
Data & 121 & 380\\\midrule
Fake-$\tauhad$ background   & 129 $\pm$ 12 & 350 $\pm$ 31 \\
$\qqbar \to ZZ$ and $gg\to ZZ$   &  $<0.01$  &  $<0.01$   \\
$H\to ZZ^*$   & $<0.01$ & 0.09 $\pm$ 0.04  \\\midrule
Total background   & 129 $\pm$ 12  &  350 $\pm$ 31   \\\midrule
$H\to aa\to 4\tau$ ($m_a = 4\,\GeV$)  & 20.2 $\pm$ 3.2 & 21.4 $\pm$ 3.3 \\
$H\to aa\to 4\tau$ ($m_a = 6\,\GeV$)  & 9.7 $\pm$ 1.5 & 10.7 $\pm$ 1.7 \\
$H\to aa\to 4\tau$ ($m_a = 8\,\GeV$)  & 7.8 $\pm$ 1.3 & 6.9 $\pm$ 1.1 \\
$H\to aa\to 4\tau$ ($m_a = 10\,\GeV$)  & 6.6 $\pm$ 1.1 & 6.0 $\pm$ 1.0  \\
$H\to aa\to 4\tau$ ($m_a = 12\,\GeV$)  & 3.7 $\pm$ 0.6 & 3.9 $\pm$ 0.6  \\
$H\to aa\to 4\tau$ ($m_a = 14\,\GeV$)  & 3.1 $\pm$ 0.5 & 2.7 $\pm$ 0.5  \\
$H\to aa\to 4\tau$ ($m_a = 15\,\GeV$)  & 2.4 $\pm$ 0.4 &  2.4 $\pm$ 0.4  \\\bottomrule
\hline\hline
\end{tabular}
\end{table}

\begin{figure}[tbp!]
\centering
\includegraphics[width=0.8\textwidth]{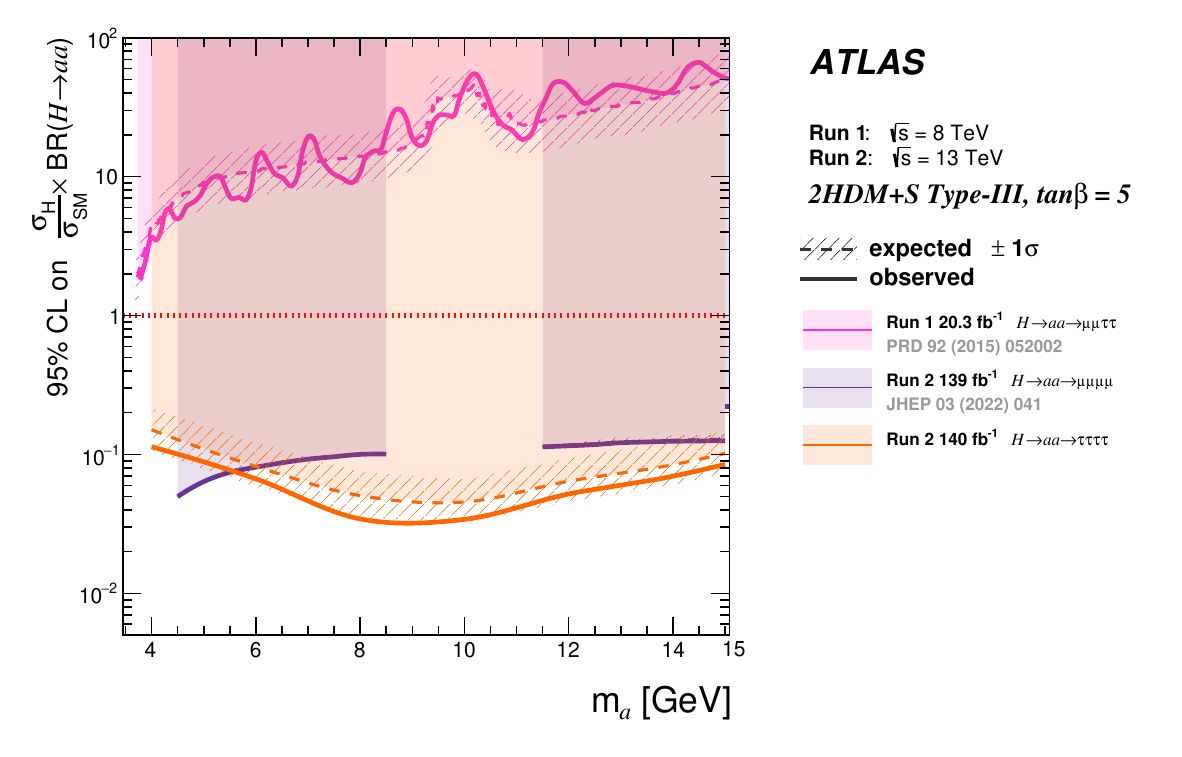}
\caption{The observed (solid line) and expected (dashed line) limits are interpreted within the framework of the Type-III 2HDM+S model with $\tan\beta=5$, which maximizes the branching ratio of $a \to \tau^+\tau^-$. These limits are expressed in terms of $(\sigma(H)/\sigma_{\text{SM}}(H))\times\mathcal{B}(H\to aa)$ as a function of $m_a$. The observed limits for the $H \to aa \to \mu\mu\mu\mu$ channel are identical to the expected limits, as no events are observed in the corresponding search. The dotted red line represents the scenario where the branching ratio of the Higgs boson exotic decay $H \to aa$ equals 100\%, assuming $\sigma(H)=\sigma_{\text{SM}}(H)$. The branching ratios of $a\to\tau^+\tau^-$ are calculated based on the methodology described in Ref.~\cite{Haisch:2018kqx}.\label{fig:HaaLimitsATLAS}}
\end{figure}

The limits presented here are comparable to other recent results in the same final state~\cite{CMS-HIG-18-006,CMS-HIG-18-024}. Once interpreted in the 2HDM+S model~\cite{Haisch:2018kqx}, as illustrated in Figure~\ref{fig:HaaLimitsATLAS}, this final state provides strong constraints for the Type-III 2HDM+S model with $\tan\beta = 5$, offering significant improvements over previous ATLAS results in the $\mu\mu\tau\tau$~\cite{HIGG-2014-02} and $\mu\mu\mu\mu$~\cite{HDBS-2018-55} final states.

\FloatBarrier


%

%

%
%
%

%
\section{Conclusion}
%

This letter presents the first ATLAS search for exotic Higgs boson decays in the $H\to aa\to \tau^+\tau^-\tau^+\tau^-$ final state, where $a$ is a new light scalar with a mass range of $4\,\GeV < m_a < 15\,\GeV$. The analysis was performed using the 140~\ifb\ of \pp collision data collected between 2015 and 2018 with the ATLAS detector at the LHC. For the first time in ATLAS, a muon-removal technique is used to reconstruct merged di-$\tau$ decays. No significant excess of data over the background expectation is observed. Upper limits at 95\% CL are set on $(\sigma(H)/\sigma_{\text{SM}}(H))\times\mathcal{B}(H\to aa\to 4\tau)$, ranging from 0.03 to 0.10 depending on $m_a$. This measurement significantly improved previous similar measurements.


%
%

%
%
%
%
%
\section*{Acknowledgements}
%

%

%
%

%
%

We thank CERN for the very successful operation of the LHC and its injectors, as well as the support staff at
CERN and at our institutions worldwide without whom ATLAS could not be operated efficiently.

The crucial computing support from all WLCG partners is acknowledged gratefully, in particular from CERN, the ATLAS Tier-1 facilities at TRIUMF/SFU (Canada), NDGF (Denmark, Norway, Sweden), CC-IN2P3 (France), KIT/GridKA (Germany), INFN-CNAF (Italy), NL-T1 (Netherlands), PIC (Spain), RAL (UK) and BNL (USA), the Tier-2 facilities worldwide and large non-WLCG resource providers. Major contributors of computing resources are listed in Ref.~\cite{ATL-SOFT-PUB-2025-001}.

We gratefully acknowledge the support of ANPCyT, Argentina; YerPhI, Armenia; ARC, Australia; BMWFW and FWF, Austria; ANAS, Azerbaijan; CNPq and FAPESP, Brazil; NSERC, NRC and CFI, Canada; CERN; ANID, Chile; CAS, MOST and NSFC, China; Minciencias, Colombia; MEYS CR, Czech Republic; DNRF and DNSRC, Denmark; IN2P3-CNRS and CEA-DRF/IRFU, France; SRNSFG, Georgia; BMFTR, HGF and MPG, Germany; GSRI, Greece; RGC and Hong Kong SAR, China; ICHEP and Academy of Sciences and Humanities, Israel; INFN, Italy; MEXT and JSPS, Japan; CNRST, Morocco; NWO, Netherlands; RCN, Norway; MNiSW, Poland; FCT, Portugal; MNE/IFA, Romania; MSTDI, Serbia; MSSR, Slovakia; ARIS and MVZI, Slovenia; DSI/NRF, South Africa; MICIU/AEI, Spain; SRC and Wallenberg Foundation, Sweden; SERI, SNSF and Cantons of Bern and Geneva, Switzerland; NSTC, Taipei; TENMAK, T\"urkiye; STFC/UKRI, United Kingdom; DOE and NSF, United States of America.

Individual groups and members have received support from BCKDF, CANARIE, CRC and DRAC, Canada; CERN-CZ, FORTE and PRIMUS, Czech Republic; COST, ERC, ERDF, Horizon 2020, ICSC-NextGenerationEU and Marie Sk{\l}odowska-Curie Actions, European Union; Investissements d'Avenir Labex, Investissements d'Avenir Idex and ANR, France; DFG and AvH Foundation, Germany; Herakleitos, Thales and Aristeia programmes co-financed by EU-ESF and the Greek NSRF, Greece; BSF-NSF and MINERVA, Israel; NCN and NAWA, Poland; La Caixa Banking Foundation, CERCA Programme Generalitat de Catalunya and PROMETEO and GenT Programmes Generalitat Valenciana, Spain; G\"{o}ran Gustafssons Stiftelse, Sweden; The Royal Society and Leverhulme Trust, United Kingdom.

In addition, individual members wish to acknowledge support from Armenia: Yerevan Physics Institute (FAPERJ); CERN: European Organization for Nuclear Research (CERN DOCT); Chile: Agencia Nacional de Investigaci\'on y Desarrollo (FONDECYT 1230812, FONDECYT 1240864); China: Chinese Ministry of Science and Technology (MOST-2023YFA1605700, MOST-2023YFA1609300), National Natural Science Foundation of China (NSFC - 12175119, NSFC 12275265); Czech Republic: Czech Science Foundation (GACR - 24-11373S), Ministry of Education Youth and Sports (ERC-CZ-LL2327, FORTE CZ.02.01.01/00/22\_008/0004632), PRIMUS Research Programme (PRIMUS/21/SCI/017); EU: H2020 European Research Council (ERC - 101002463); European Union: European Research Council (BARD No. 101116429, ERC - 948254, ERC 101089007), European Regional Development Fund (SMASH COFUND 101081355, SLO ERDF), Horizon 2020 Framework Programme (MUCCA - CHIST-ERA-19-XAI-00), European Union, Future Artificial Intelligence Research (FAIR-NextGenerationEU PE00000013), Horizon 2020 (EuroHPC - EHPC-DEV-2024D11-051), Italian Center for High Performance Computing, Big Data and Quantum Computing (ICSC, NextGenerationEU); France: Agence Nationale de la Recherche (ANR-21-CE31-0013, ANR-21-CE31-0022, ANR-22-EDIR-0002); Germany: Baden-Württemberg Stiftung (BW Stiftung-Postdoc Eliteprogramme), Deutsche Forschungsgemeinschaft (DFG - 469666862, DFG - CR 312/5-2); China: Research Grants Council (GRF); Italy: Istituto Nazionale di Fisica Nucleare (ICSC, NextGenerationEU), Ministero dell'Università e della Ricerca (NextGenEU 153D23001490006 M4C2.1.1, NextGenEU I53D23000820006 M4C2.1.1, NextGenEU I53D23001490006 M4C2.1.1); Japan: Japan Society for the Promotion of Science (JSPS KAKENHI JP22H01227, JSPS KAKENHI JP22H04944, JSPS KAKENHI JP22KK0227, JSPS KAKENHI JP23KK0245, JSPS KAKENHI JP24K23939); Norway: Research Council of Norway (RCN-314472); Poland: Ministry of Science and Higher Education (IDUB AGH, POB8, D4 no 9722), Polish National Science Centre (NCN 2021/42/E/ST2/00350, NCN OPUS 2023/51/B/ST2/02507, NCN OPUS nr 2022/47/B/ST2/03059, NCN UMO-2019/34/E/ST2/00393, UMO-2020/37/B/ST2/01043, UMO-2022/47/O/ST2/00148, UMO-2023/49/B/ST2/04085, UMO-2023/51/B/ST2/00920, UMO-2024/53/N/ST2/00869); Portugal: Foundation for Science and Technology (FCT); Spain: Generalitat Valenciana (Artemisa, FEDER, IDIFEDER/2018/048), Ministry of Science and Innovation (MCIN \& NextGenEU PCI2022-135018-2, MICIN \& FEDER PID2021-125273NB, RYC2019-028510-I, RYC2020-030254-I, RYC2021-031273-I, RYC2022-038164-I); Sweden: Carl Trygger Foundation (Carl Trygger Foundation CTS 22:2312), Swedish Research Council (Swedish Research Council 2023-04654, VR 2021-03651, VR 2022-03845, VR 2022-04683, VR 2023-03403, VR 2024-05451), Knut and Alice Wallenberg Foundation (KAW 2018.0458, KAW 2022.0358, KAW 2023.0366); Switzerland: Swiss National Science Foundation (SNSF - PCEFP2\_194658); United Kingdom: Leverhulme Trust (Leverhulme Trust RPG-2020-004), Royal Society (NIF-R1-231091); United States of America: U.S. Department of Energy (ECA DE-AC02-76SF00515), Neubauer Family Foundation.

%
%


%
\clearpage
\printbibliography
\clearpage
 
\begin{flushleft}
\hypersetup{urlcolor=black}
{\Large The ATLAS Collaboration}

\bigskip

\AtlasOrcid[0000-0002-6665-4934]{G.~Aad}$^\textrm{\scriptsize 104}$,
\AtlasOrcid[0000-0001-7616-1554]{E.~Aakvaag}$^\textrm{\scriptsize 17}$,
\AtlasOrcid[0000-0002-5888-2734]{B.~Abbott}$^\textrm{\scriptsize 123}$,
\AtlasOrcid[0000-0002-0287-5869]{S.~Abdelhameed}$^\textrm{\scriptsize 119a}$,
\AtlasOrcid[0000-0002-1002-1652]{K.~Abeling}$^\textrm{\scriptsize 55}$,
\AtlasOrcid[0000-0001-5763-2760]{N.J.~Abicht}$^\textrm{\scriptsize 49}$,
\AtlasOrcid[0000-0002-8496-9294]{S.H.~Abidi}$^\textrm{\scriptsize 30}$,
\AtlasOrcid[0009-0003-6578-220X]{M.~Aboelela}$^\textrm{\scriptsize 45}$,
\AtlasOrcid[0000-0002-9987-2292]{A.~Aboulhorma}$^\textrm{\scriptsize 36e}$,
\AtlasOrcid[0000-0001-5329-6640]{H.~Abramowicz}$^\textrm{\scriptsize 157}$
% ...
\bigskip
\\

% ...
$^{17}$Department for Physics and Technology, University of Bergen, Bergen; Norway.\\
% ...
$^{30}$Physics Department, Brookhaven National Laboratory, Upton NY; United States of America.\\
% ...
$^{36}$$^{(a)}$Facult\'e des Sciences Ain Chock, Universit\'e Hassan II de Casablanca;$^{(b)}$Facult\'{e} des Sciences, Universit\'{e} Ibn-Tofail, K\'{e}nitra;$^{(c)}$Facult\'e des Sciences Semlalia, Universit\'e Cadi Ayyad, LPHEA-Marrakech;$^{(d)}$LPMR, Facult\'e des Sciences, Universit\'e Mohamed Premier, Oujda;$^{(e)}$Facult\'e des sciences, Universit\'e Mohammed V, Rabat;$^{(f)}$Institute of Applied Physics, Mohammed VI Polytechnic University, Ben Guerir; Morocco.\\
% ...
$^{45}$Physics Department, Southern Methodist University, Dallas TX; United States of America.\\
% ...
$^{49}$Fakult\"{a}t Physik , Technische Universit{\"a}t Dortmund, Dortmund; Germany.\\
% ...
$^{55}$II. Physikalisches Institut, Georg-August-Universit\"{a}t G\"ottingen, G\"ottingen; Germany.\\
% ...
$^{104}$CPPM, Aix-Marseille Universit\'e, CNRS/IN2P3, Marseille; France.\\
% ...
$^{119}$$^{(a)}$New York University Abu Dhabi, Abu Dhabi;$^{(b)}$United Arab Emirates University, Al Ain; United Arab Emirates.\\
% ...
$^{123}$Homer L. Dodge Department of Physics and Astronomy, University of Oklahoma, Norman OK; United States of America.\\
% ...
$^{157}$Raymond and Beverly Sackler School of Physics and Astronomy, Tel Aviv University, Tel Aviv; Israel.\\
% ...
$^{a}$ Also at Affiliated with an institute formerly covered by a cooperation agreement with CERN.\\
$^{b}$ Also at An-Najah National University, Nablus; Palestine.\\
$^{c}$ Also at Borough of Manhattan Community College, City University of New York, New York NY; United States of America.\\
$^{d}$ Also at Center for Interdisciplinary Research and Innovation (CIRI-AUTH), Thessaloniki; Greece.\\
$^{e}$ Also at Centre of Physics of the Universities of Minho and Porto (CF-UM-UP); Portugal.\\
$^{f}$ Also at CERN, Geneva; Switzerland.\\
% ...

\end{flushleft}


%

%
%

%

%
%
%
%

\end{document}